\documentstyle[bo99,epsfig]{article}

\title{Resolving the 10-40 keV Cosmic X-ray Background with Constellation-X}
\author{Giorgio Matt, Fulvio Pompilio \& Fabio La Franca }
\affil{Dipartimento di Fisica,
 Universit\`a degli Studi Roma Tre,
via della Vasca Navale 84, I--00146 Roma, Italy
 }

\begin{document}

\maketitle

\begin{abstract}

The energy density of the Cosmic X--ray background (XRB) peaks around 30 keV
(see Figure 1), 
an energy
not yet probed by focussing imaging instruments. 
The first hard X--ray telescope due to fly on a space mission
will be that on board Constellation--X. The
imaging capability, besides providing an improvement of several orders
of magnitude in sensitivity over current passively collimated detectors, will
permit for the first time to resolve a fraction of the XRB at this most crucial 
energy. Synthesis models of the XRB based on obscured AGN predict that at
least 40\% of the 10--40 keV XRB will be resolved by Constellation--X.

\keywords{X--rays: galaxies; galaxies: nuclei }
\end{abstract}

\section{ The Constellation X--ray mission }

The Constellation X--ray mission is a high throughput X-ray
facility emphasizing observations at high spectral resolution
(E/$\Delta$=300--3000) while covering a broad energy bandpass
(0.25--40 keV). Constellation-X will provide a factor of nearly
100 increase in sensitivity over current high resolution X-ray
spectroscopy missions. The large collecting area is achieved with a
design utilizing several mirror modules, each with its own detector system.
Each (or a few) science unit will fly on a separate spacecraft.

Two telescopes will be on--board: the low--energy Spectroscopy
X-ray telescope (SXT), operating simultaneously with
a 2 eV resolution calorimeter and a set of reflection gratings; a high--energy
system (HXT), that will be the first focusing telescope system operating
at several tens of keV, where the energy density of the XRB
peaks (Figure 1).

\medskip
The Baseline Mission Characteristics are:

\begin{itemize}

\item Effective Area: 15000 (6000, 1500) cm$^2$ at 1 (6.4, 40) keV

\item Angular resolution: 15'' HPD from 0.25 to 10 keV; 1' HPD at 40 keV

\item Band Pass: 0.25 to 40 keV

\end{itemize}

\noindent
More information can be found at:~~~{\sc http://constellation.gsfc.nasa.gov/}

\section{ The XRB synthesis model}

To evaluate the fraction of hard XRB resolved by the HXT onboard
Constellation-X, we
first developed a synthesis model based on the standard assumption that
the XRB is mostly made by a combination of type 1 and 2 AGN (Setti \& Woltjer
1989; Comastri
et al., 1995, and references therein). Details on the model
can be found in Pompilio, La Franca \& Matt (1999 and this volume). 
Here we summarize the main features of the model.

\begin{enumerate}
\item AGN spectra
\begin{enumerate}
\item type 1 (AGN1) spectrum:
\begin{itemize}
\item power law ($\alpha=0.9$) + exponential cut-off (${E}_{c}=400$ keV);
\item Compton reflection component (accretion disk, ${\theta }_{obs}\sim
{60}^{\circ }$);
\end{itemize}
\item type 2 (AGN2) spectrum (Matt, Pompilio \& La Franca, 1999):
\begin{itemize}
\item primary AGN1 spectrum obscured by cold matter:\par ${10}^{21}\le
{N}_{H}\le {10}^{25}{cm}^{-2}$, $\frac{dN(log{N}_{H})}{d(log{N}_{H})}\propto
log{N}_{H}$;
\item Compton scattering within the absorbing matter fully included.
\end{itemize}
\end{enumerate}

\item Cosmological evolution
\begin{enumerate}
\item PLE (${\Phi }^{*}(z=0)=1.45\times {10}^{-6}
{Mpc}^{-3}{({10}^{44}erg~{s}^{-1})}^{-1}$);
\item power law evolution for the break-luminosity:\par
 ${L}^{*}(z)\propto
{(1+z)}^{k}$ up to ${z}_{max}=1.73$, with \par
${L}^{*}(z=0)=3.9\times {10}^{43}
erg~{s}^{-1}$ and $k=2.9$ (model H of Boyle et al., 1994);
\item the redshift integration is performed up to ${z}_{d}=4.5$.
\end{enumerate}
\end{enumerate}

%--------------------------  figure 1
%this section shows how to insert a figure in the text
\begin{figure}
\centerline{\psfig{file=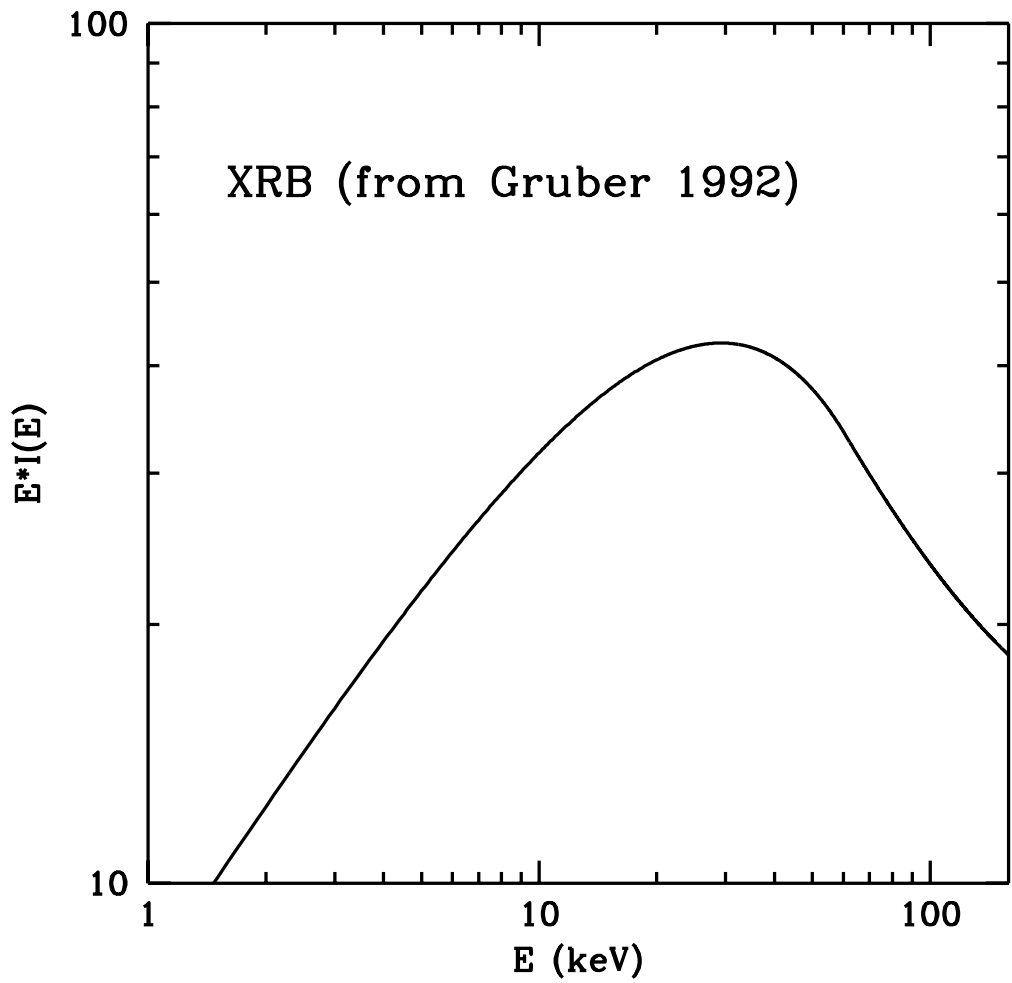, width=15cm}}
\caption[]{The spectrum of the Cosmic X--ray Background after Gruber (1992).
The $\nu F_{\nu}$ representation shows that the energy density peaks at 30 keV.}
\end{figure}
%---------------------------------

\section{ Predictions}

We are now able to predict the fraction of the Cosmic XRB which can be resolved
by Constellation--X in the 10--40 keV energy range. Our estimate is
based on the baseline spatial resolution, i.e. 1' HPD, 
which is a rather conservative value. Any improvement in this
resolution will of course increase the fraction of XRB resolved. 

\bigskip
The predictions are:

\begin{enumerate}
\item Number densities of type 1 and type 2
AGN detected in the 10-40 keV band down to a flux limit of
${10}^{-14}~erg~{cm}^{-2}{s}^{-1}$ (reachable with an exposure time of
the order of a few thousand seconds, Harrison et al. 1999):
\begin{itemize}
\item AGN1 $\longrightarrow 53.3~{deg}^{-2}$;
\item AGN2 $\longrightarrow 123.4~{deg}^{-2}$;
\end{itemize}
$\Rightarrow$  $\sim $ 70\% are absorbed sources.
(Note that these numbers implied a density of about a source per 20 beams,
which should ensure a tolerable level of confusion. If the final spatial
resolution will be better than 1', as it appears likely, confusion
problems will be of course less severe). 
\item Integrated XRB spectrum in the 10-40 keV band:
\begin{itemize}
\item measured value $\longrightarrow {I}(10-40~keV)\simeq 9.4\times
{10}^{-8}~erg~{cm}^{-2}{s}^{-1}{sr}^{-1}$ (HEAO-1 A2 data, Marshall et al.
1980);
%\item model value $\longrightarrow {I}_{mod}(10-40~keV)\simeq 8.7\times
%{10}^{-8}~erg~{cm}^{-2}{s}^{-1}{sr}^{-1}$;
\item fraction of XRB down to a flux of ${10}^{-14}~erg~{cm}^{-2}
{s}^{-1}$ \par $\longrightarrow {I}_{Const-X}(10-40~keV)\simeq 4.0\times
{10}^{-8}~erg~{cm}^{-2}{s}^{-1}{sr}^{-1}$;
\end{itemize}

\bigskip
$\Rightarrow$
{\bf $\sim $ 40\% of the 10--40 keV XRB will
be resolved by Constellation-X.}

\end{enumerate}

\begin{acknowledgements}
This work is part of the activities of the Constellation--X Facility Science
Team. The authors acknowledge financial support from the Italian Space Agency.
\end{acknowledgements}

\end{document}